\documentclass[a4paper,fleqn]{cas-sc}

\usepackage[authoryear]{natbib}
\usepackage{cleveref}
\usepackage{placeins}
\usepackage{siunitx}
\usepackage{rotating}

\AtBeginDocument{%
}

\def\tsc#1{\csdef{#1}{\textsc{\lowercase{#1}}\xspace}}
\tsc{WGM}
\tsc{QE}
\ExplSyntaxOn
\cs_gset:Npn \__first_footerline:
{
    \group_begin:
    \small
    \sffamily
    \__short_authors: :~
    { \rmfamily \itshape Preprint~---~Not~peer~reviewed~---~Version~1,~October~2026 }
    \group_end:
}
\ExplSyntaxOff

\begin{document}
\let\WriteBookmarks\relax

\shorttitle{Sounds Cheesy}    

\shortauthors{W. Feng \emph{et al.}}  

\title [mode = title]{Sounds Cheesy - Bridging the Texture Gap Between Dairy and Plant-Based cheese with Non-Contact Ultrasound and AI}  




\author[1]{Wenjing Feng}%[<options>]

\author[1,2]{Thomas A. Hazlehurst}%[orcid=0000-0001-9767-4391]

\author[1,2]{Nik Watson}[orcid=0000-0001-5216-4873]
                       % linkedin=nik-watson-4312a63a]
\cormark[1]

% % Footnote of the second author

% Email id of the second author
\ead{n.j.watson@leeds.ac.uk}

% % URL of the second author

\date{\today}


\affiliation[1]{organization={School of Food Science and Nutrition, University of Leeds},
            city={Leeds},
%          citysep={}, % Uncomment if no comma needed between city and postcode
            postcode={LS2 9JR}, 
            country={UK}}

\affiliation[2]{organization={National Alternative Protein Innovation Centre},
            % city={},
%          citysep={}, % Uncomment if no comma needed between city and postcode
            country={UK}}

\cortext[1]{Corresponding author}



\begin{abstract}
Plant-based cheeses often fail to reproduce the texture of dairy cheese. Closing this gap is slow because the reference methods, texture profile analysis (TPA) and sensory panels, destroy the sample and are poorly suited to screening many formulations in a timely and cost-effective manner. Here we show that non-contact, air-coupled ultrasonic imaging combined with machine learning can characterise cheese texture rapidly and without damaging the sample. Thirteen commercial products (nine dairy, four plant-based; 119 slices, 357 scans) were imaged in through-transmission at 100 kHz to give two-channel amplitude and time-of-flight C-scans, and each slice was then measured by TPA. The C-scans revealed millimetre-scale acoustic domains in ripened dairy cheeses, consistent with their fused-curd microstructure, that were absent from the homogeneous starch and oil matrix of the plant-based products. TPA showed that the texture gap is one of range and failure behaviour rather than average firmness. Mean hardness was similar for the two categories, but dairy products spanned a 7.5-fold range against 1.9-fold for plant-based products, and the plant-based products failed in a brittle, strain-softening manner rather than by ductile yielding. No single ultrasonic parameter tracked texture across products. However, a compact convolutional neural network predicted full true stress–strain loading and unloading curves from the C-scan images of unseen slices across all thirteen products without product-specific calibration. Clustering the predicted curves produced a texture map on which three plant-based slices sat with the Cheddars and one with Emmental, while soft, mozzarella-like textures remained unoccupied. The approach offers a one-minute, non-contact measurement suited to formulation screening and at-line quality control of cheese.\end{abstract}


% % Research highlights
% \begin{highlights}
% \item 
% \item 
% \item 
% \end{highlights}

% Keywords

\maketitle

% Main text
\section{Introduction}\label{introduction}

Cheese is one of the most widely consumed dairy foods in the world, valued for its nutritional density and its distinctive sensory character. Among its sensory attributes, texture is central: the firmness, springiness and melting behaviour of cheese strongly influence consumer perception, acceptance and repeat purchase, \cite{Szczesniak2002Texture,Foegeding2003CheeseTexture}. These properties emerge from a complex, heterogeneous microstructure including a ripened casein network filled with fat globules and moisture which evolves with composition, manufacture and maturation, \cite{Fox2017CheeseStructure,Lamichhane2018CheeseStructureFunction}. At the same time, there is growing pressure to move away from animal-derived foods. Dairy cheese carries one of the largest environmental footprints in the food sector, with greenhouse gas emissions of approximately 24 kg CO$_2$-equivalent per kilogram of product, compared with less than 2.5 kg for the ingredients used in typical plant-based alternatives, \cite{Ritchie2022EnvironmentalImpacts}. Combined with concerns over land and water use and animal welfare, and dietary requirements such as lactose intolerance, milk-protein allergy and vegan lifestyles, \cite{Jeske2018PlantBasedDairy,Viroli2023PlantBasedDiets}, these drivers have created strong consumer and industrial momentum behind plant-based cheese.

Plant-based cheese is one of the fastest-growing plant-based food categories, with retail sales in the United States alone increasing by more than 70\% between 2018 and 2020, \cite{GFI2021RetailSales}. Most commercial products are formulated from starches, coconut oil and hydrocolloids, structured as a gel, and typically contain little or no protein, \cite{Fresan2021PlantBasedCheeseNutrition,Grasso2021PlantBasedBlocks}. This composition produces a fundamentally different structure from dairy cheese: a homogeneous starch--oil matrix rather than a heterogeneous, fat-filled casein network, \cite{Grossmann2021PlantBasedCheese,Dobson2025StructuralDifferences}. As a consequence, current plant-based cheeses reproduce neither the wide range of textures found across dairy varieties nor their characteristic mechanical and melting behaviour, \cite{Dobson2023PlantBasedCheese}, and consumers consistently identify texture as a principal barrier to acceptance, \cite{Short2021PlantBasedCheeses,Falkeisen2022PlantBasedCheese,Ali2025PlantBasedCheese}.

Closing this texture gap is currently slow and expensive. Developing a plant-based formulation that mimics a dairy target means iterating over protein sources, starches, fats and process conditions, with each candidate assessed by destructive instrumental methods including texture profile analysis (TPA), rheology, melt and stretch tests and ultimately by trained sensory panels, \cite{Bourne2002FoodTexture,Foegeding2007CheeseTexture,Dobson2023PlantBasedCheese}. These reference methods are validated and remain indispensable, but they are labour-intensive, destroy the sample, characterise only bulk properties and are ill-suited to screening large numbers of formulations. There is therefore a clear need for measurement technologies that can assess cheese texture rapidly, non-destructively and at scale to both to benchmark plant-based prototypes against dairy targets during development, and to monitor product consistency during manufacture.

Low-intensity ultrasonic sensing is a strong candidate because the propagation of an ultrasonic wave is governed by the elastic properties, density and microstructure of the medium it travels through. Ultrasonic measurements have long been used to characterise food materials non-destructively, \cite{Mizrach2008UltrasonicFruit,Arvanitoyannis2017Ultrasounds,Villamiel2017UltrasoundFoodProcessing}. In cheese specifically, ultrasonic velocity and attenuation have been related to composition and maturity in Cheddar, \cite{Benedito2000CheddarUltrasound}, and used to monitor the ripening of Torta del Casar \cite{Crespo2020TortaDelCasar}. Most of this work used contact or immersion configurations, which require a coupling gel or liquid; for food this risks contaminating the product and deforming its surface, and it is poorly suited to sliced or ready-to-eat formats and to production environments, \cite{Cho2003CheeseUltrasound}. Non-contact, air-coupled ultrasound removes the couplant entirely and has been applied to a range of food inspection problems, including imaging and monitoring the coagulation of milk-based products, \cite{Gan2006NonContactUltrasound}. \cite{Cho2003CheeseUltrasound} showed that air-coupled measurements could determine the thickness of hard cheeses with sub-0.1 mm accuracy, and that ultrasonic velocity correlated strongly ($R^2 > 0.9$) with Young's modulus, hardness and toughness across five varieties. Baek et al. \cite{Baek2012CheeseUltrasound} subsequently used the same air-instability-compensation sensors to estimate TPA texture parameters including fracturability, hardness, springiness and chewiness of the same five hard cheeses from ultrasonic velocity, with regression models reaching $R^2$ values of up to 0.95. Furthermore, by raster-scanning the transducers across the sample, air-coupled systems can produce C-scan images that map amplitude and time-of-flight over the sample area, capturing the spatial heterogeneity that single-point acoustic measurements and destructive bulk tests such as TPA cannot. Low-intensity ultrasound has also begun to be applied to plant-based foods, with air-coupled measurements recently used to track the mechanical properties of high-moisture meat analogues during and after production, \cite{Guillermic2023MeatAnalogues,Rangira2026PeaProteinUltrasound}; plant-based cheese, however, has not yet been examined.

The obstacle is no longer acquiring ultrasonic data on samples but interpreting it. Ultrasonic parameters respond simultaneously to composition, microstructure and geometry, and simple scalar features correlate with mechanical properties inconsistently: attenuation showed no significant relationship with mechanical properties in the study of Cho and Irudayaraj, \cite{Cho2003CheeseUltrasound}, and remained a markedly weaker predictor than velocity in later work, \cite{Baek2012CheeseUltrasound}. Ultrasonic velocity--texture relationships established for one variety of cheese generally do not transfer to others, so each product requires its own calibration, \cite{Benedito2000CheddarUltrasound,Crespo2020TortaDelCasar}. Averaging a C-scan image into a single number discards precisely the spatial information that distinguishes a heterogeneous ripened cheese from a homogeneous plant-based gel.

Machine learning (ML), a common type of Artificial Intelligence, addresses this bottleneck by learning the relationship between rich sensor data and target properties directly from examples, without requiring an explicit physical model. ML is increasingly applied across cheese production, from milk quality through ripening to final product assessment, \cite{Pardo2025MachineLearningCheese}, and optical spectral measurements combined with neural networks have been used to model cheese hardness during ripening, \cite{Vasquez2018CheeseHardness}. However, it is important to note that optical methods sense composition rather than the mechanical response itself, whereas ultrasound, as a mechanical wave, directly probes the elastic behaviour that underlies texture, \cite{Prempeh2025UltrasoundFood}. Ultrasonic measurements combined with ML have already been used to monitor fermentation processes, \cite{Bowler2021BeerUltrasound}, the removal of fouling during clean-in-place operations, \cite{Escrig2020FoulingUltrasound}, and the mixing of ingredients, \cite{Bowler2020MixingUltrasound}, and intelligent sensors of this kind are increasingly viewed as enabling technologies for sustainable, digitalised food manufacturing, \cite{Watson2021IntelligentSensors}. Convolutional neural networks (CNNs) are particularly well suited to spatially structured, image-like data, \cite{Krizhevsky2012ImageNet}, making them a natural match for ultrasonic C-scan images: they can learn the local acoustic patterns associated with microstructure and map them to mechanical behaviour. This enables the prediction of not just a single texture scalar, but the full instrumented texture response, as recently demonstrated for the stress--strain curves of engineered composites from microstructure images, \cite{Yang2020CompositeStressStrain}, but not yet from ultrasonic images of food.

In this work we couple non-contact, air-coupled ultrasonic C-scan imaging with machine learning to characterise the texture of thirteen commercial cheeses, nine dairy and four plant-based. Specifically, we (i) train a compact CNN to predict TPA true stress and strains curves (loading and unloading, directly from two-channel (ultrasonic amplitude and velocity) C-scan images, providing a non-contact, non-destructive estimate of instrumented texture; and (ii) cluster the cheeses based on the TPA and ML results.

To our knowledge, this is the first study to predict full instrumented texture curves from ultrasonic C-scan images using deep learning, and the first to apply non-contact ultrasound and machine learning to compare the texture of dairy and plant-based cheeses. The approach is intended as a rapid screening tool: a way to place any cheese (including a new plant-based prototypes) on a texture map alongside established dairy products and thereby accelerate the development of plant-based cheeses that close the texture gap.

\section{Materials and Methods}\label{mat&met}

\subsection{Food Samples}\label{food-samples}

Thirteen commercial cheese products were selected to span a wide range of textures across both dairy and plant-based categories \cref{cheese-table}. The nine dairy products comprised mature Cheddar, 30\% reduced-fat Cheddar, extra mature Cheddar, hard and semi-hard Gouda, mozzarella, Emmental, Red Leicester and smoked Cheddar. The four plant-based products comprised two coconut-oil-based slices (a Cheddar alternative and a smoked mature flavour), Cathedral City plant-based slices and Violife Original slices. All products were purchased as pre-cut slices from supermarkets in Leeds, UK (Tesco, Morrisons and Marks \& Spencer) between late June and early July 2025.

A total of 119 slices were tested (85 dairy, 34 plant-based; per-product numbers in \cref{cheese-table}). Each slice was labelled with cheese type, pack number, slice number and measurement date to ensure traceability. After purchase, products were stored in sealed plastic bags at $4^{\circ}\mathrm{C}$ until testing. The thickness of each slice was measured and mean $\pm$ SD thickness per product is given in Table~1. Slices were removed from refrigerated storage immediately before each measurement, so that all products entered testing from the same $4^{\circ}\mathrm{C}$ starting condition; ambient conditions during the measurements are given in \cref{exp-meas}.

\begin{table}[h] 
\centering
\begin{tabular}{llp{0.15\linewidth}lp{0.25\linewidth}ll}
\toprule
Code   & Category & Product name             & Brand & Ingredients       & No. of slices & Thickness \\ \midrule
A01-CH & Dairy    & Mature Cheddar           & Tesco & Milk              & 10            &  $2.90 \pm 0.14$       \\
A02-RE & Dairy    & 30\% Reduced Fat Cheddar & Tesco & Milk (Irish milk) & 10            &  $2.92 \pm 0.07$        \\
A03-EX & Dairy    & Extra Mature Cheddar     & Tesco & Milk (UK milk)    & 10            &  $2.92 \pm 0.05$        
\\
A04-GD  & Dairy	 & Gouda (hard)   & Tesco	& Milk (Dutch milk)	& 10 & $2.46 \pm 0.31$
\\
A05-GD	& Dairy	& Gouda (semi-hard) &	Tesco	& Milk, Salt, Stabiliser: Calcium Chloride; Starter Culture	& 8 & $1.94 \pm 0.20$
\\
A06-MO	& Dairy	& Mozzarella	& Morrisons	& Milk (British \& EU Milk) &	10	& $2.84 \pm 0.41$
\\
A07-EM	& Dairy	& Emmental &	Morrisons &	Milk (Dutch milk)	& 10	& $2.79 \pm 0.23$
\\
A08-LE	& Dairy &	Red Leicester &	Morrisons &	Red Leicester cheese(cheese (milk), colour (annatto norbisin)) &	10 & $2.78 \pm 0.12$
\\
A09-SM	& Dairy &	Smoked Cheddar & Applewood	& Dairy casein, fat, smoked flavour	& 7 & $2.83 \pm 0.17$
\\
P01-SM	& Plant	 & Smoked mature flavour (coconut oil-based) &	M\&S &	Water, Coconut Oil (22\%), Modified Potato Starch, Smoked Water, Cornflour, Salt, Modified Corn Starch, Calcium, Yeast Extract, Thickener: Carrageenan, Guar Gum, Acidity Regulator: Lactic Acid, E325, Flavourings, Rowan Extract, Tapioca Starch, Smoked Paprika, Colour: Carotenes. &	9	& $2.56 \pm 0.01$
\\
P02-CH	& Plant &	Cheddar alternative (coconut oil-based)	& M\&S &	Water, Coconut Oil (22\%), Modified Potato Starch, Cornflour, Salt, Modified Maize Starch, Thickener: Carrageenan, Guar Gum, Tricalcium Citrate, Yeast Extract, Flavourings, Acidity Regulator: Lactic Acid, E325, Colour: Carotenes. &	9 & $2.57 \pm 0.06$
\\
P03-CC &	Plant &	Cathedral City Plant-based	& Cathedral city	& Rice base, coconut oil(26\%), modified starch, Potato starch, Sea salt, Flavourings, Calcium, Bamboo Fibre, Acid, Gelling Agent, Colour &	6 & $2.36 \pm 0.08$
\\
P04-VO	& Plant &	Violife Original &	Violife &	Water, Modified starch, Coconut oil(23\%), Sea salt, Rowanberry extract, Flavourings, Acidity regulator, Olive extract, Colour, Vitamin B12 &	10	& $2.11 \pm 0.08$
\\
\bottomrule
\end{tabular}
\caption{Note: “A” indicates dairy-based products and “P” indicates plant-based products. The number denotes product order (01–09 for dairy, 01–04 for plant-based), and the suffix (e.g., -CH, -RE, -VO) is an abbreviation of the product name.}\label{cheese-table}
\end{table}

\subsection{Experimental measurement}\label{exp-meas}
\subsubsection{Non-contact ultrasound}
Non-contact ultrasonic measurements were performed with an Ultran U170 air-coupled ultrasonic scanning system (The Ultran Group, Boalsburg, PA, USA) operating in through-transmission mode. Two matched air-coupled transducers (Ultran NCG100-D25-P76; nominal centre frequency 100~kHz, 25~mm active element diameter, point-focused with a focal length of 76~mm) were mounted on opposite arms of the scanning gantry, one as transmitter and one as receiver, with a fixed separation of 125~mm. Each slice rested on a transparent plate with a 40~mm diameter central hole to minimise boundary reflections, with the cheese sample directly between the two transducers.

The transmitter converts the electrical excitation into an airborne ultrasonic wave, which propagates through the air gap and the cheese slice and is detected by the receiver. The excitation was a five-cycle burst at 85~kHz and 390~V, with a pulse repetition frequency of 100~Hz and positive polarity. The receiver gain was 50~dB, with 80~kHz high-pass and 800~kHz low-pass filters. Data acquisition was controlled with SecondWave Research Studio 5.1, and full waveforms were saved (\texttt{*.ipp} format) for offline processing.

Scans were acquired at three positions within the central region of every slice, each covering approximately $20 \times 20$~mm with a step size of $\approx 0.26$~mm; one scan took approximately one minute. In total, 357 scans were acquired from the 119 slices (three per slice). Slices were scanned immediately after removal from $4^{\circ}\mathrm{C}$ storage and returned to storage afterwards for texture analysis, \cref{text-anal}. Scanning took place at an ambient laboratory temperature of $23 \pm 2^{\circ}\mathrm{C}$.

\begin{figure}[pos=!htbp]
  \centering
    \includegraphics[width=\linewidth,height=0.58\textheight,keepaspectratio]{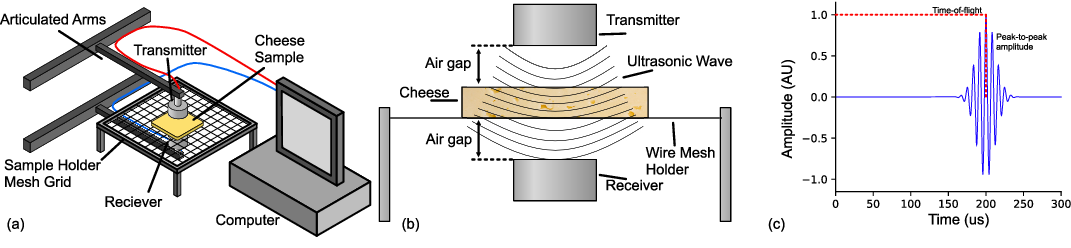}
    \caption{Experimental set up of (a) the \emph{Ultran U710x with NCG100-D25-P76 transducers}, (b) the ultrasonic set up with received waveform and acoustic path, (c) the peak-to-peak and time of flight measurements on a received wave}\label{fig1}
\end{figure}

From each waveform, two quantities were extracted within a fixed time gate around the first transmitted wave packet (\Cref{fig1}): the peak-to-peak amplitude, expressed in mV and in dB, and the time of flight (ToF), taken as arrival time of the packet peak, measured relative to the excitation trigger. Mapping these values across the scan area produced two-dimensional C-scan images of amplitude and ToF. Peak-to-peak amplitudes were normalised by slice thickness (mV/mm) to account for differences in acoustic path length through the samples.

An apparent propagation velocity was calculated as slice thickness divided by the measured ToF. Because the measured ToF is dominated by the fixed 125~mm air path, this quantity is not the material sound speed of the cheese; it is used only as a relative index for comparing products of similar geometry.

Amplitude attenuation and propagation velocity are sensitive to density, moisture content and structural heterogeneity, \cite{Mizrach2008UltrasonicFruit,Benedito2000CheddarUltrasound,Crespo2020TortaDelCasar}, while the air-coupled configuration requires no coupling medium, preserving sample integrity and allowing rapid, non-destructive measurement of fragile food slices. The experimental setup, acoustic path and example received waveform are displayed in \cref{fig1}. 

%%%% FIGURE 1

\subsubsection{Texture Analysis}\label{text-anal}
Texture profile analysis (TPA) was performed with a \emph{TA.XT2i} texture analyser (\emph{Stable Micro Systems}, Godalming, UK) fitted with a \SI{40}{\milli\metre} flat-faced cylindrical probe (P/40). In TPA the sample is compressed twice to a set strain to mimic the first two bites of chewing, and texture parameters are derived from the resulting force--time curve; these correlate with sensory texture perception in cheese and other semi-solid foods, \cite{Bourne2002FoodTexture,Szczesniak2002Texture}.

After ultrasonic scanning, cylindrical sub-samples ($\SI{10}{\milli\metre}$ diameter) were cut from the scanned regions of each slice using a cutting tool and stacked to a height of approximately \SI{7}{\milli\metre} to form one test specimen per slice. One measurement was made per slice, giving 119 TPA measurements matching the 119 scanned slices.

Specimens were compressed to \SI{50}{\percent} strain at a test speed of $\SI{1}{\milli\metre\per\second}$ (pre-test $\SI{2}{\milli\metre\per\second}$, post-test $\SI{5}{\milli\metre\per\second}$), with a $\SI{5.0}{\gram}$ trigger force. Measurements were carried out under controlled laboratory conditions ($\SI{23 \pm 2}{\degreeCelsius}$), and force and displacement were calibrated before testing following the manufacturer's procedure.

From each force--time curve the standard TPA parameters were calculated, \cite{Bourne2002FoodTexture}: hardness (peak force during the first compression), springiness ($L_2/L_1$, the extent to which the specimen recovers its height between compressions), cohesiveness ($A_2/A_1$, the ratio of the positive area under the second compression to that under the first), adhesiveness (the negative force area as the probe withdraws after the first compression), resilience (the ratio of the withdrawal area to the compression area within the first cycle) and chewiness (hardness $\times$ cohesiveness $\times$ springiness).

Hardness was selected as the principal reference parameter because it is robust and widely used as an instrumental indicator of cheese firmness in instrumental and sensory studies, \cite{Zheng2016CheeseTexture,Cho2003CheeseUltrasound}. Although destructive, TPA remains a widely used and validated standard for cheese texture and provides the reference measurements for developing and evaluating the non-destructive ultrasonic approach. For the machine-learning analysis (\Cref{sect-ml}), the full force--displacement records were exported in addition to the scalar TPA parameters for the first compression--decompression cycle only.

\subsection{Machine Learning Methods}\label{sect-ml}

The repository for the machine-learning work can be found at [GitHub link TBD]. The code was written in \emph{Python} 3.12 and used \emph{PyTorch} 2.10.

\subsubsection{Prediction of stress--strain curves}

\textbf{Dataset.} Each observation paired a two-channel C-scan image with the mechanical response of the slice it came from. The two channels were the thickness-normalised attenuation map and the ToF map, each $\numproduct{76 x 76}$ pixels, giving inputs of shape $(2, 76, 76)$. Because one TPA measurement was made per slice, the three scans of a slice share that slice's measured curve, and all observations from the same slice were treated as a single group throughout model evaluation. The final dataset contained 348 scan--curve pairs from 116 slices.

\textbf{Targets.} Measured force--displacement curves were converted to true stress--true strain using the specimen geometry via the conversion for true strain
\begin{equation}
    \epsilon = -\ln \left(1 - \frac{\Delta L}{L_0}\right),
\end{equation}
and true stress
\begin{equation}
    \sigma = F \frac{\left(1 - \frac{\Delta L}{L_0}\right)}{A_0},
\end{equation}
assuming constant volume and uniform deformation, where $\Delta L$ is the change in distance, $L_0$ is the initial distance, and $A_0$ is the initial area. Each curve was split into loading and unloading branches, and each branch was interpolated to 32 fixed points on a common normalised strain axis, giving targets of shape $(2, 32)$. This places curves of differing native sampling density into a consistent fixed-length format, so the task becomes fixed-length curve regression: the model maps spatial texture information to the full mechanical response rather than to a single scalar property. Targets were standardised using the mean and standard deviation of the training set only, preventing information from the validation or test sets influencing training.

\textbf{Architecture.} A compact convolutional neural network (CNN) was used, chosen because the inputs are spatially structured, image-like measurements: convolutional layers capture local texture features, spatial gradients and repeated patterns with far fewer parameters than a fully connected network applied to all pixels, and a small architecture limits overfitting given the modest number of independent groups. The network comprises three convolutional blocks with $2 \rightarrow 16 \rightarrow 32 \rightarrow 64$ channels and $3 \times 3$ kernels, each using group normalisation, chosen over batch normalisation for robustness to small batch sizes, and GELU activations, with max-pooling after the first two blocks. Adaptive average pooling condenses the final 64-channel feature maps into a 64-dimensional feature vector by averaging each channel over its spatial dimensions. A fully connected regression head then maps this vector through a 128-unit hidden layer to 64 outputs, reshaped into the two 32-point branches.

\textbf{Training.} Inputs were lightly augmented with small spatial shifts of up to 3 pixels, per-channel scale jitter of $\pm$\qty{5}{\percent}, and additive Gaussian noise with a standard deviation of \num{0.01}, encouraging the model to learn robust texture--response relationships rather than memorising exact spatial arrangements. The model was trained with the AdamW optimiser with weight-decay regularisation using a learning rate of \num{0.001}, weight decay of \num{0.001}, batch size of 16, a maximum of 400 epochs, and early-stopping patience of 50 epochs. The loss combined a Smooth $L_1$ term on the predicted curve points with a smoothness penalty on adjacent points, weighted by \num{0.05}: the former rewards accurate prediction of the measured values, while the latter favours physically plausible curve shapes. Early stopping was applied on validation loss and the best checkpoint retained. Hyperparameters were selected manually based on validation performance, and random seeds were fixed to 42 for the data splitting and model training.

\begin{figure}[pos=!htbp]
  \centering
    \includegraphics[width=\linewidth,height=0.58\textheight,keepaspectratio]{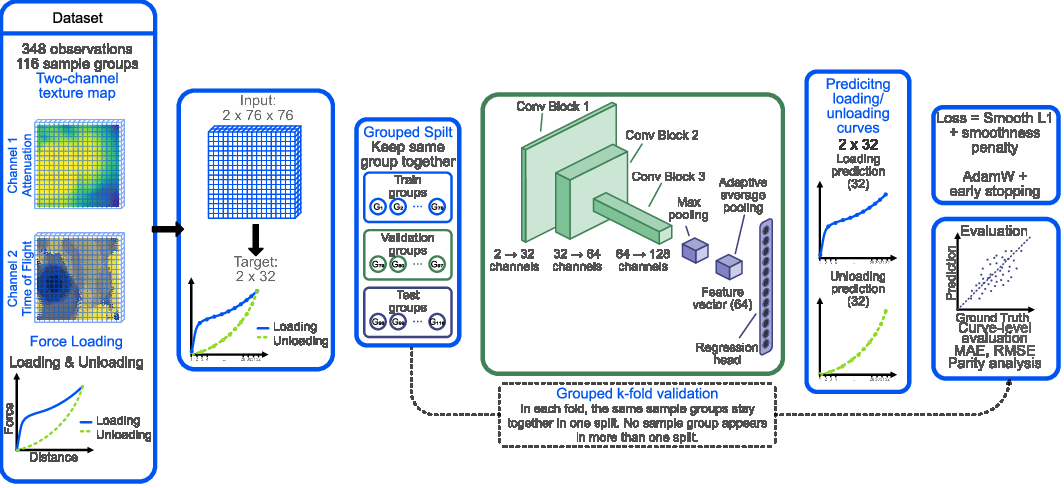}
    \caption{A flow diagram of the machine learning network used to predict the force-distance curves trained on the ultrasonic image data}\label{fig2}
\end{figure}

\textbf{Evaluation.} Performance was estimated using grouped 5-fold cross-validation, with an additional \qty{10}{\percent} grouped held-out test set. Groups corresponded to individual slices, ensuring that replicate scans from the same slice were never split across training, validation or test partitions. This prevents leakage between repeated measurements of the same physical slice and provides a conservative estimate of generalisation to unseen slices from known product classes. Model outputs were transformed back from the normalised training scale to the original force scale before computing performance metrics, including mean absolute error, root-mean-square error and parity analysis ($R^2$) between predicted and measured curves, reported overall and per product. Predictions shown in the results section correspond to held-out test slices that were not used for model fitting or validation. Generalisation to entirely unseen products was not assessed in the present study and is reserved for future leave-one-product-out evaluation. An overview of the modelling methodology is available in \cref{fig2}.

%%%% FIGURE 2

\subsubsection{Texture clustering}

To assess whether cheese samples grouped according to texture, PCA and clustering were applied separately to three feature representations: measured force curves, raw ultrasound images, and neural-network-derived image features. For the force-based analysis, each measured loading/unloading force curve was flattened into a single feature vector and used as the texture reference. For the ultrasonic image analysis, the two-channel ultrasound images were flattened and analysed directly. For the neural-network analysis, each image was passed through the trained CNN, and the resulting 64-dimensional feature vector from the final convolutional feature block was used. In all three cases, features were standardised feature-wise before PCA. Clustering was then performed with k-means on the first 10 principal components, while the first two principal components were used for visualisation. The number of clusters was selected by silhouette analysis over $k = 2$ to 8, unless a fixed k was specified. Agreement between clustering methods was quantified using the adjusted Rand index, including comparisons between the force-curve clusters and the raw-image or CNN-feature clusters. The force-curve clustering therefore served as the texture-based reference against which the ultrasound-derived representations were compared.

\section{Results \& Discussion}\label{sec:results}

\subsection{Experimental Results}\label{sec:expresults}

\begin{figure}[pos=!htbp]
  \centering
    \includegraphics[width=\linewidth,height=0.58\textheight,keepaspectratio]{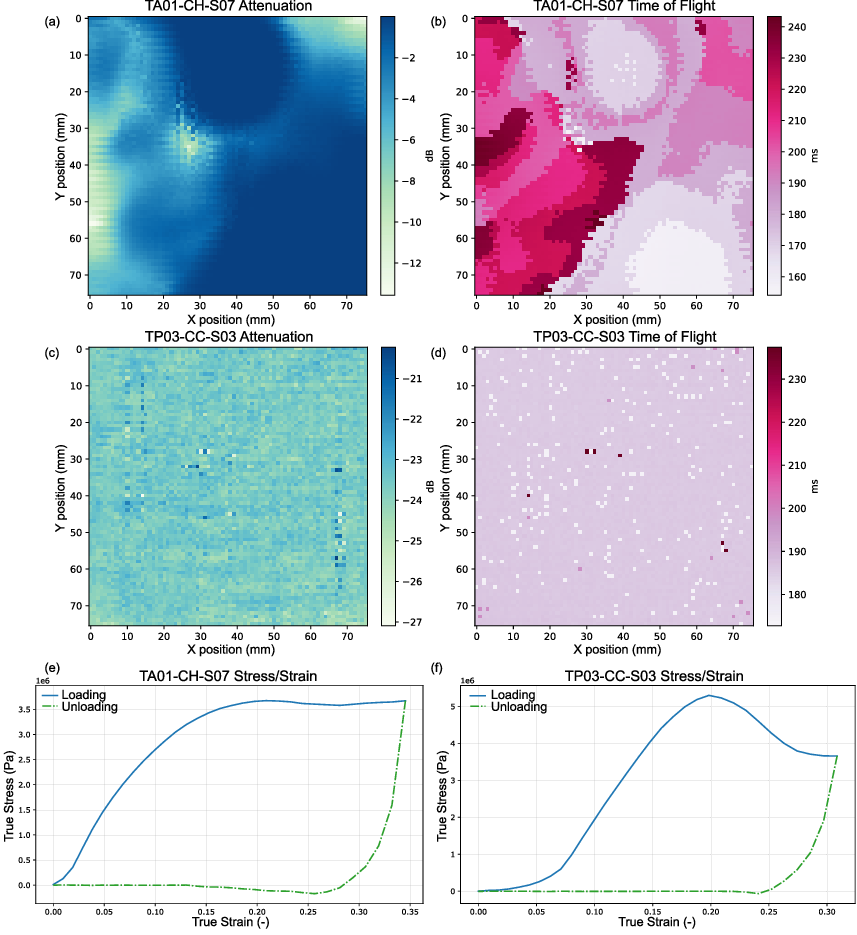}
    \caption{Non-contact ultrasonic C-scan images and texture profile analysis of a mature Cheddar slice (A01-CH, slice S07; panels a, b, e) and a Cathedral City plant-based slice (P03-CC, slice S03; panels c, d, f). (a, c) Transmitted amplitude (\si{\decibel}) and (b, d) time of flight (\si{\micro\second}) over a \qtyproduct{20 x 20}{\milli\metre} scan area ($76 \times 76$ measurement points); (e, f) the corresponding true stress--true strain curves from the first compression cycle (loading, solid line; unloading, dashed line). Note that the two attenuation maps are drawn on different colour scales.}\label{fig3} 
    % add labels to individual images
\end{figure}

\Cref{fig3} shows representative ultrasonic C-scan images and the corresponding TPA stress--strain curves for a mature Cheddar slice (A01-CH, slice S07) and a Cathedral City plant-based slice (P03-CC, slice S03), the two cheeses with the greatest visual contrast in the dataset. The two products differ in two separate respects. The first is spatial structure. The attenuation map of the Cheddar slice (\Cref{fig3}(a)) is organised into contiguous acoustic domains several millimetres across, within which the transmitted amplitude varies by roughly \qty{10}{\decibel} (from about \qty{-2}{\decibel} to \qty{-12}{\decibel}) inside a single \qtyproduct{20 x 20}{\milli\metre} scan, and the same domain pattern appears in the time-of-flight map (\Cref{fig3}(b)) as shifts of several tens of microseconds. The plant-based slice shows no comparable structure: both of its maps (\Cref{fig3}(c), (d)) are a fine, structureless speckle whose pixel-to-pixel variation is a few decibels and about \qty{10}{\micro\second}, with only isolated outlying pixels. The second difference is overall transmission. The plant-based slice attenuated the transmitted wave by a further \qtyrange{15}{20}{\decibel} (its colour scale runs from \qty{-21}{\decibel} to \qty{-27}{\decibel}, against \qty{-2}{\decibel} to \qty{-12}{\decibel} for the Cheddar), to the extent that the two maps cannot usefully be displayed on a common scale. A single one-minute non-contact scan therefore separates this pair of dairy and plant-based products on two independent grounds before any modelling is applied.

%%%% FIGURE 3

The domain structure in the Cheddar maps is what would be expected from the microstructure of a ripened, pressed curd. Cheddar consists of fused curd particles whose junctions are richer in protein and poorer in fat than their interiors, and whose fat, moisture and mineral distributions vary locally with manufacture and maturation, \cite{Fox2017CheeseStructure,Lamichhane2018CheeseStructureFunction}. Each of these features changes the local density and elastic modulus, and therefore the local acoustic impedance, so that the transmitted amplitude and arrival time vary from one region of the slice to the next. Synchrotron X-ray microtomography of commercial Cheddar shows exactly this organisation, with curd boundaries outlined by mineral crystals and protein-rich junctions within an otherwise uniform fat-filled casein matrix, whereas commercial plant-based cheese appears as a homogeneous continuous starch matrix with finely and evenly dispersed coconut-oil droplets and no larger-scale organisation, \cite{Dobson2025StructuralDifferences}. The C-scans of \Cref{fig4} reproduce this structural distinction non-destructively. The origin of the much stronger attenuation of the plant-based product is less certain: Table~2 shows that the plant products were among the most attenuating in the study, but that the two Goudas (A04, A05) attenuated as strongly, so attenuation level alone does not separate the categories. Plausible contributions are greater intrinsic absorption in a water-rich starch and hydrocolloid gel and differences in surface finish and thickness between sliced products.

A second, coarser source of heterogeneity in the dairy products was visible to the eye. Many of the Cheddar and Gouda slices contained small slit-like mechanical openings, and the Emmental slices contained large round eyes characteristic of that variety, whereas the plant-based slices were free of openings. Scan positions were chosen to avoid visible openings where possible, but slits narrower than the scan area, openings just beneath the cut surface and the margins of larger eyes cannot be excluded from a \qtyproduct{20 x 20}{\milli\metre} scan. The effect of an opening on an airborne through-transmitted wave depends on whether it passes through the full thickness of the slice or is enclosed within it. A through-opening offers the wave an unobstructed air path, so it appears as a localised region of anomalously high transmitted amplitude, with an arrival time shifted by the absence of the cheese path; an enclosed void introduces two additional cheese-to-air interfaces at which nearly all of the incident energy is reflected, so it appears as a region of very low amplitude. In either case the feature is superimposed on the finer domain structure of the curd, and its size is set by the opening rather than by the microstructure. It is not possible to tell from the maps alone whether any of the features in \cref{fig3}a are of this kind. The openings are a real property of these products rather than an artefact of the measurement, but they add a source of acoustic variation that is only loosely related to the mechanical response of the surrounding material.

The stress--strain curves in \cref{fig3}(e) and (f) show that the acoustic contrast is not simply a proxy for firmness. The mature Cheddar loads to a plateau of roughly \qty{3.7}{\mega\pascal} that is sustained from about \num{0.2} to \num{0.35} true strain, so the casein and fat network yields but continues to carry load, and it recovers very little of its deformation on unloading, the stress returning to zero after only a small reduction in strain. This is a ductile, largely plastic response. The Cathedral City slice reaches a higher peak stress (about \qty{5.3}{\mega\pascal} at \num{0.2} strain), then strain-softens by about a third over the remainder of the compression, and its unloading branch also collapses almost immediately, which is the signature of a stiff, brittle gel that fractures internally rather than deforming. In firmness terms this plant product is harder than the mature Cheddar (\qty{1947 \pm 125}{\gram} against \qty{1540 \pm 187}{\gram}; \cref{results-table}), so the difference that the ultrasound detects is one of structure and failure mode, not of strength. This is consistent with texture profile analysis of commercial dairy and plant-based cheese in the literature: reported hardness values of commercial plant-based slices lie on both sides of dairy Cheddar, \cite{Dobson2023PlantBasedCheese}, but their springiness, cohesiveness and chewiness are consistently lower than those of dairy cheese, which has been attributed to the dense, highly organised retrograded-starch matrix of these products, \cite{Dobson2025StructuralDifferences}.

Across the full set of products (\cref{results-table}), dairy firmness spanned a 7.5-fold range, from \qty{590 \pm 65}{\gram} for the smoked Cheddar slices to \qty{4421 \pm 363}{\gram} for Emmental, with mozzarella and hard Gouda below \qty{1000}{\gram} and Red Leicester and reduced-fat Cheddar above \qty{3000}{\gram}. The four plant-based products occupied a narrow central band of \qtyrange{1462}{2837}{\gram}, a 1.9-fold range: the two coconut-oil slices (P01, P02) closely matched the mature and extra-mature Cheddars (A01, A03) in firmness, Cathedral City was somewhat firmer, and Violife (\qty{2837 \pm 122}{\gram}) approached the hard dairy products. None of the plant products reached either the soft end of the dairy range (mozzarella, smoked Cheddar, hard Gouda) or its hard end (Emmental). The plant products were also more uniform. Their within-product coefficients of variation in firmness were \qtyrange{4}{6}{\percent}, against \qtyrange{6}{14}{\percent} for the dairy products, and in thickness-normalised amplitude \qtyrange{1}{6}{\percent} for P01--P03 (\qty{23}{\percent} for P04) against \qtyrange{8}{66}{\percent} for dairy. This is the difference between an industrially standardised gel and a biological raw material that varies with source and ripening, and for the acoustic measurements it also reflects the absence from the plant-based slices of both the curd structure and the openings described above. Mean firmness was in fact similar for the two categories (approximately \qty{2050}{\gram} for dairy and \qty{1950}{\gram} for plant-based), so the texture gap seen here is one of range and failure behaviour rather than of average firmness. This is a more precise statement of the gap than ``plant-based cheese is softer'', and one that matches the wide range of hardness over which a single plant-based formulation can be tuned, \cite{Dobson2023PlantBasedCheese}.

\begin{table}[h] 
\begin{tabular}{p{0.25\linewidth}lllll}
\toprule
Cheese ID &	Group &	Firmness ($\unit{\gram}$) &	Pk2Pk ($\unit[per-mode = symbol]{\milli\volt\per\milli\metre}$) &	Pk2Pk ($\unit{\decibel}$) & Velocity ($\unit[per-mode = symbol]{\milli\metre\per\micro\second}$) \\
\midrule
A01 Cheddar	& Dairy & $1539.64 \pm 187.42$ & $266.76 \pm 106.36$ & $-4.26 \pm 1.60$ & $0.0141 \pm 0.0009$ \\
A02 30\% Reduced Fat &	Dairy &	$3437.69 \pm 270.88$ & $50.80 \pm 17.13$ & $-8.25 \pm 0.54$	& $0.0153 \pm 0.0004$ \\
A03 Extra Mature Cheddar &	Dairy & $1582.94 \pm 156.08$ & $256.07 \pm 76.83 $ &  $-6.18 \pm 1.37 $ & $0.0140 \pm 0.0005$ \\
A04 Hard Gouda & Dairy & $991.93 \pm 134.81$ & $59.21 \pm 15.47$ & $-9.59 \pm 0.93$ & $0.0133 \pm 0.0009$ \\
A05 Semi-hard Gouda	& Dairy & $1788.88 \pm 156.87$ & $110.02 \pm 27.23$ & $-10.74 \pm 1.19$ & $0.0101 \pm 0.0011$ \\
A06 Mozzarella & Dairy & $933.95 \pm 53.94$ & $66.22 \pm 43.65$ & $-8.37 \pm 1.09$ & $0.0148 \pm 0.0023$ \\
A07 Emmental & Dairy & $4420.72 \pm 362.78$ & $75.34 \pm 37.29$ & $-7.85 \pm 0.82$ & $0.0144 \pm 0.0015$ \\
A08 Red Leicester & Dairy & $3163.73 \pm 399.75$ & $42.63 \pm 3.53$ & $-8.85 \pm 0.37$ & $0.0150 \pm 0.0007$ \\
A09 Smoke flavoured cheddar & Dairy & $589.95 \pm 64.69$ & $58.64 \pm 9.64$ & $-7.78 \pm 0.32$ & $0.0152 \pm 0.0009$ \\
P01 Coconut oil based smoked flavour & Plant & $1543.54 \pm 67.22$ & $52.61 \pm 0.48$ & $-9.19 \pm 0.06$ & $0.0138 \pm 0.0012$ \\
P02 Coconut oil based mature Cheddar & Plant & $1462.23 \pm 86.19$ & $51.60 \pm 2.70$ & $-9.22 \pm 0.23$ & $0.0138 \pm 0.0004$ \\
P03 Cathedral City & Plant & $1947.43 \pm 124.66$ & $56.18 \pm 3.08$ & $-10.01 \pm 0.31$ & $0.0127 \pm 0.0005$ \\
P04 Violife	& Plant & $2836.78 \pm 121.58$ & $81.11 \pm 18.31$ & $-10.33 \pm 0.56$ & $0.0112 \pm 0.0005$ \\
\bottomrule
\end{tabular}
\caption{Mean $\pm$ standard deviation of firmness and ultrasound parameters (peak-to-peak amplitude in $\unit[per-mode = symbol]{\milli\volt\per\milli\metre}$ and $\unit{\decibel}$, and velocity) for 13 cheese types.}\label{results-table}
\end{table}

The ultrasonic parameters in \cref{results-table} show a different ordering. In thickness-normalised amplitude the two mature Cheddars stood apart from everything else, transmitting \qty{267 \pm 106}{\milli\volt\per\milli\metre} and \qty{256 \pm 77}{\milli\volt\per\milli\metre}, against \qtyrange{43}{110}{\milli\volt\per\milli\metre} for the other dairy products and \qtyrange{52}{81}{\milli\volt\per\milli\metre} for the plant products, between two and six times more. In decibel terms the plant products were the most attenuating and most tightly grouped category (\qtyrange{-9.2}{-10.3}{\decibel}, with standard deviations of \qtyrange{0.06}{0.56}{\decibel}), while the dairy values spread from \qty{-4.3}{\decibel} (A01) to \qty{-10.7}{\decibel} (A05). The apparent velocity varied comparatively little between products (\qtyrange{0.010}{0.015}{\milli\metre\per\micro\second}) and its ranking followed neither firmness nor category. Within products, Spearman correlations between firmness and any single ultrasonic parameter were mostly weak and non-significant; the exceptions were the smoked Cheddar (dB: $\rho = \num{0.86}$, $p = \num{0.014}$; velocity: $\rho = \num{0.82}$, $p = \num{0.023}$) and Violife (mV/mm: $\rho = \num{-0.69}$, $p = \num{0.029}$; velocity: $\rho = \num{0.70}$, $p = \num{0.025}$).

These weak scalar relationships contrast with the strong velocity--hardness calibrations reported for five hard cheeses measured with \qty{1}{\mega\hertz} air-coupled transducers ($R^2 > \num{0.9}$; \cite{Cho2003CheeseUltrasound,Baek2012CheeseUltrasound}), but they are consistent with two other findings of that work: that attenuation, being sensitive to surface unevenness, minor defects, porosity and uneven component distribution, showed no significant relationship with mechanical properties, and that ultrasound--texture relationships established for one variety do not transfer to others, so that each product requires its own calibration \cite{Benedito2000CheddarUltrasound,Crespo2020TortaDelCasar}. The slit openings and eyes noted above are exactly the kind of defect and porosity that make a scalar attenuation value unreliable. Over a set that spans soft, eyed, smoked, reduced-fat and starch-based products, no single scalar acoustic parameter calibrates for texture. The point of \cref{fig3}, however, is that the sensitivity of attenuation to the distribution of components, a nuisance for a scalar measurement, is precisely what makes the attenuation \emph{image} informative: the domain structure is a direct map of that distribution, and it is what distinguishes a heterogeneous ripened cheese from a homogeneous gel of the same firmness. An image also allows a feature that is unrelated to texture, such as an opening, to be recognised by its size and contrast rather than averaged into the measurement. Retaining the full two-channel image, rather than reducing it to its mean, is therefore the natural route to a texture model that does not need a separate calibration for every product, and it is the motivation for the convolutional approach evaluated in the next section.

\subsection{Predicting stress–strain curves from ultrasonic images}\label{sec:pred}

\begin{figure}[pos=!htbp]
  \centering
    \includegraphics[width=\linewidth,height=0.58\textheight,keepaspectratio]{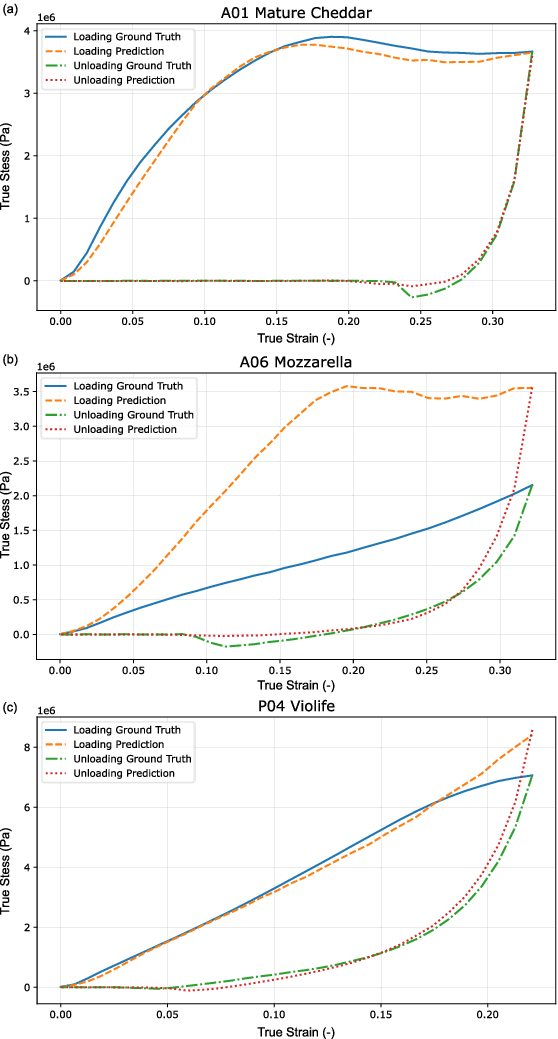}
    \caption{Predicted and measured loading and unloading stress–strain curves for three held-out slices: (a) mature Cheddar (A01), (b) mozzarella (A06), (c) Violife (P04). Predictions are those of the grouped cross-validation fold in which each slice was in the test set. Panels (a) and (c) illustrate good agreement for a dairy and a plant-based product; panel (b) shows the largest systematic error observed in the study.}\label{fig4}
\end{figure}

\Cref{fig4} compares predicted and measured loading and unloading curves for three slices held out from training, chosen to span the range of model behaviour: a mature Cheddar (A01), a mozzarella (A06) and a Violife slice (P04). For the mature Cheddar (\cref{fig4} a) the predicted loading branch reproduces the form of the measured curve, with an initial rise, a maximum of about \qty{3.8}{\mega\pascal} near \num{0.17} strain and a gently declining plateau, while underestimating the stress by \qtyrange{0.2}{0.4}{\mega\pascal} at low and intermediate strains; the unloading branch, including the strain at which the stress returns to zero, is reproduced almost exactly. For Violife (\cref{fig4} c), the stiffest product in the study, both branches are tracked over the full strain range. The predicted loading curve lies within about \qty{5}{\percent} of the measured one up to \num{0.17} strain and then overshoots the measured maximum (\qty{7.1}{\mega\pascal}) by roughly \qty{20}{\percent} at the end of compression, and the gradual, elastic unloading branch, which shows that this product recovers most of its deformation, in contrast to the Cheddar, is reproduced with a slight overestimate at intermediate strains. The mozzarella slice (\cref{fig4} b) is the failure case. The model overestimates the loading stress by about \qty{65}{\percent} at maximum strain (\qty{3.5}{\mega\pascal} against \qty{2.2}{\mega\pascal}) and by more than a factor of two at intermediate strain and, more tellingly, assigns the sample the wrong shape: a Cheddar-like rise to a plateau in place of the measured gradual, almost linear increase in stress. The network has, in effect, applied the response of a generic semi-hard cheese to a sample whose acoustic signature it could not distinguish from those products.

%%%% FIGURE 4

\begin{figure}[pos=!htbp]
  \centering
    \includegraphics[width=\linewidth,height=0.58\textheight,keepaspectratio]{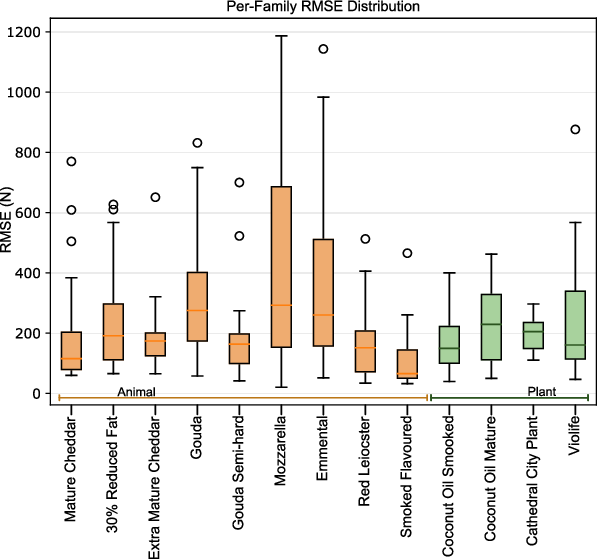}
    \caption{Distribution of the root-mean-square error between predicted and measured stress–strain curves for each product for the nine dairy (orange) and four plant-based (green) products. Boxes show the interquartile range and median, whiskers extend to 1.5 × IQR and circles mark outliers.}\label{fig5}
\end{figure}

\Cref{fig5} summarises the curve-level RMSE for every held-out prediction, grouped by product. The four plant-based products were predicted with moderate and consistent error: medians of roughly \num{150} (P01), \num{230} (P02), \num{205} (P03) and \num{165} (P04), interquartile ranges of similar width and, apart from a single Violife outlier, no extreme values. The errors for the dairy products depended strongly on the product. The smoked Cheddar slices (median $\approx \num{65}$) and mature Cheddar ($\approx \num{115}$) were predicted best, followed by Red Leicester, semi-hard Gouda, extra-mature Cheddar and reduced-fat Cheddar (medians \numrange{150}{190}), whereas mozzarella (median $\approx \num{290}$, upper quartile $\approx \num{690}$, maximum $\approx \num{1190}$), hard Gouda ($\approx \num{275}$, maximum $\approx \num{830}$) and Emmental ($\approx \num{260}$, maximum $\approx \num{1140}$) were predicted worst, with tails extending to four to five times their medians. Read together, \cref{fig4,fig5} tell one story: mozzarella is both the illustrative failure and the widest error distribution.

%%%% FIGURE 5

The products that were hard to predict are neither the softest nor the hardest but those whose mechanical behaviour is unusual relative to their acoustic signature. This can be read directly from the clustering figure discussed in the next section. In the projection of the ultrasonic images (\cref{fig6} a) the mozzarella, Emmental and Gouda scans lie inside the same dense cluster as the reduced-fat Cheddar, Red Leicester and the plant products, so their acoustic signatures are not distinctive, yet their measured curves lie at the extremes of the dairy range (mozzarella and hard Gouda the softest, Emmental the stiffest; \cref{results-table} and \cref{fig6} b). A model trained by minimising an average loss over a dataset in which Cheddar-type products predominate will, for a scan it cannot place, predict a curve close to the dataset average, which is exactly the plateau-shaped, intermediate-stiffness curve produced for the mozzarella slice in \cref{fig4} b, and it will therefore underestimate the stiffest and overestimate the softest products. Structural factors compound this. Mozzarella showed the largest relative variability of any product in both amplitude (CV \qty{66}{\percent}) and apparent velocity (CV \qty{16}{\percent}; \cref{results-table}), consistent with its fibrous, anisotropic pasta-filata structure and high moisture content, \cite{Fox2017CheeseStructure}. The openings described in \cref{sec:expresults} act in the same direction for the eyed and slit-bearing products. The eyes of Emmental are far larger than any other structural feature in the dataset and larger than the \qty{10}{\milli\metre} sub-sample used for TPA, and the slits in the Goudas and Cheddars are irregular in size and position, so for these products the scanned region and the compressed sub-sample need not contain the same structure: the acoustic image may record an opening that the reference measurement does not, or the reverse. This weakens the relationship that the network has to learn for exactly those products, and it is a plausible contributor to the wide error distributions of Emmental and hard Gouda in \cref{fig5}. It is not the whole explanation, since the mature Cheddars also contained slits and were predicted well. Their acoustic signatures are the most distinctive in the dataset (\cref{fig6} a), so the network can identify them and reproduce their characteristic plateau curves despite the added variability. The plant products, which are acoustically indistinctive but mechanically close to the dataset average and free of openings, are predicted with moderate error precisely because a prediction close to the average costs them little.

Predicting the full loading and unloading response, rather than a single hardness value, extends the earlier cheese and ultrasound literature in two ways. Previous studies regressed scalar mechanical properties on scalar velocity for a small number of hard cheeses of similar structure, with a separate calibration in each study, \cite{Benedito2000CheddarUltrasound,Cho2003CheeseUltrasound,Baek2012CheeseUltrasound}. Here a single model covers thirteen products of widely differing structure, including four starch-based gels, with no product-specific calibration, and returns the whole curve, from which hardness (peak stress), the presence of a yield plateau or of strain-softening, and the fraction of deformation recovered on unloading can all be read. The approach parallels the use of convolutional networks to predict the stress--strain curves of engineered composites from images of their microstructure, \cite{Yang2020CompositeStressStrain}, but uses as its image a non-contact acoustic measurement of a real food. It also differs from previous ultrasound and machine-learning work in food manufacturing, which has related waveform or scalar acoustic features to process states such as fermentation, mixing and fouling,  \cite{Bowler2020MixingUltrasound,Bowler2021BeerUltrasound,Escrig2020FoulingUltrasound}, in mapping a two-dimensional acoustic image to a mechanical response curve. Its limitations should be stated plainly. The grouped cross-validation estimates performance on unseen slices of known products, not on unseen products; the modest number of independent slices (116) limits the size of network that can be trained without overfitting; and the Smooth-$L_1$ objective with a smoothness penalty, while producing physically plausible curves, favours the majority curve shape when the input is ambiguous, as the mozzarella example shows.

\subsection{Clustering Results}\label{sec:cluster_results}

\begin{sidewaysfigure}
% \begin{figure}%[]
  \centering
  \includegraphics[width=\textwidth]{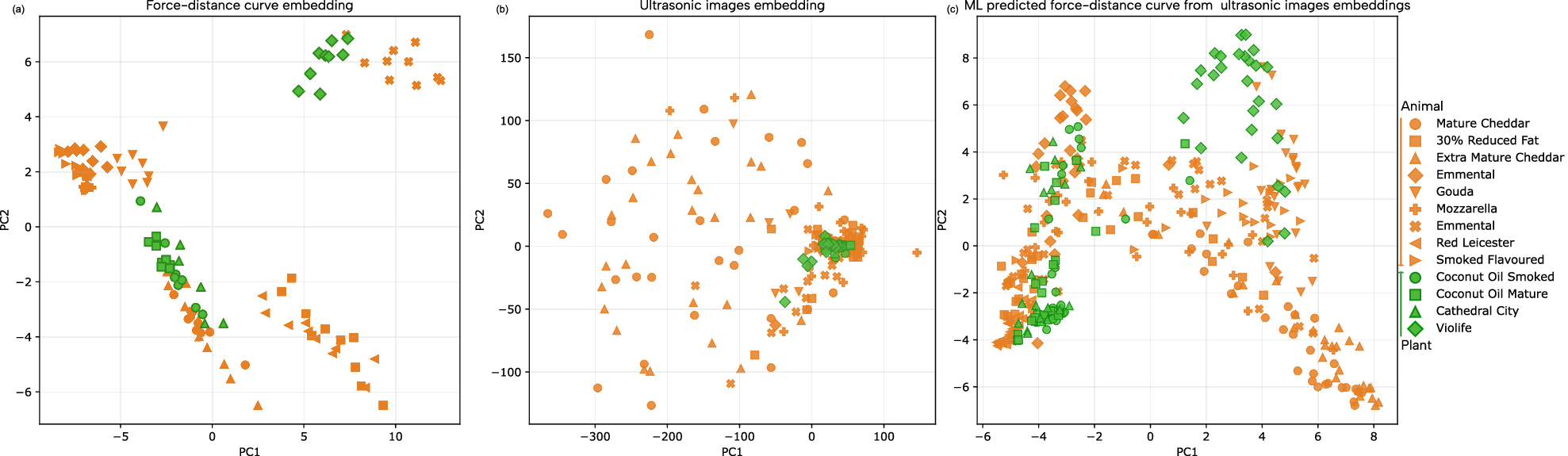}
    % \makebox[0.5\textwidth][c]{\includegraphics[width=\textwidth]{figure6.eps}}
    \caption{Two-dimensional principal-component projections of (a) the measured stress–strain curves (one point per slice), (b) the ultrasonic C-scan images, and (c) the stress–strain curves predicted from the ultrasonic images (one point per scan), for the nine dairy (orange) and four plant-based (green) products.}\label{fig6}
% \end{figure}
\end{sidewaysfigure}

\Cref{fig6} shows two-dimensional principal-component projections of three representations of the same thirteen products: (a) the measured stress--strain curves from the TPA, one point per slice; (b) the ultrasonic C-scan images; and (c) the stress--strain curves predicted by the network from the ultrasonic images, one point per scan. Panel (a) is therefore the texture ground truth against which any non-contact map must be judged; panel (b) shows how the products organise on the raw non-contact measurement alone; and panel (c) is what the non-contact route actually delivers once the learnt mapping is applied.

Quantitative clustering metrics supported the visual differences observed in the PCA projections. Clustering of the CNN-derived ultrasound embeddings identified seven clusters with a mean silhouette score of \num{0.405} and modest agreement with product family (ARI = \num{0.205}). Clustering of the raw ultrasound images produced two clusters with a higher silhouette score (\num{0.669}), but showed little agreement with product family (ARI = \num{0.039}), indicating that this separation was not strongly product-specific. By contrast, clustering of the measured force/stress--strain curves produced four clusters with a silhouette score of \num{0.440} and stronger agreement with product family (ARI = \num{0.341}). Agreement between the ultrasound embedding clusters and force/stress--strain clusters was limited but greater than for raw images (ARI = \num{0.168} versus \num{0.009}), suggesting that the learned ultrasound representation captured some, but not all, of the texture-related structure present in the mechanical measurements.

In the projection of the ultrasonic images (\cref{fig6} b) the dominant source of variance is the two mature Cheddars. Their scans are spread over a wide region (PC1 from about \num{-370} to \num{0}), while every other product, dairy and plant alike, collapses into a single dense cluster near the origin; the only other outliers are individual mozzarella and Emmental scans. The plant products lie inside this dense cluster, indistinguishable from reduced-fat Cheddar, Red Leicester, the Goudas or the smoked Cheddar, with only a few Violife scans slightly displaced. This is the picture that (\Cref{results-table}) predicts. The mature Cheddars transmit several times more energy than any other product and have the largest within-product amplitude variability, to which scans that included slit openings may have contributed, so an unsupervised decomposition of the raw images is dominated by the transmission level and heterogeneity of those two products and shows no organisation by texture or by category. The conclusion drawn for scalar parameters in \cref{sec:expresults} therefore holds for the images as well: clustering the acoustic data directly does not produce a texture classification.

The projection of the measured curves (\cref{fig6} a) organises the products largely by firmness along PC1 and by curve shape along PC2. The soft dairy products (smoked Cheddar, mozzarella and hard Gouda, with semi-hard Gouda adjacent) form a group on the left; the two hard Cheddar-type products, reduced-fat Cheddar and Red Leicester, form a group at the lower right; and Emmental forms a group at the upper right. The plant products fall in two distinct places. The three starch and coconut-oil slices (P01--P03) lie along a central band where they overlap with the mature and extra-mature Cheddars, the products whose firmness they match (\cref{results-table}). Violife, by contrast, lies at the upper right with Emmental rather than with the reduced-fat Cheddar and Red Leicester that are closer to it in firmness, indicating that the shape of its curve, an almost linear rise to high stress without a plateau followed by substantial elastic recovery (\cref{fig4} c), resembles the response of Emmental more than that of the Cheddars. In texture terms, then, each commercial plant-based slice in this study has an identifiable dairy counterpart, Cheddar-like for the three starch and coconut-oil slices and Emmental-like for Violife, and none of them approaches the soft dairy products.

The projection of the predicted curves (\cref{fig6} c) recovers the principal features of this texture map from the ultrasonic images alone. The three starch and coconut-oil slices form a single compact cluster at one extreme of PC1, adjacent to but distinguishable from a dense group of soft and intermediate dairy scans; Violife forms a separate compact cluster at the opposite corner of the map, with only a handful of dairy scans nearby; and the replicate scans of each slice fall close together. The plant products are thus resolved as two groups of Cheddar-like and Emmental-like texture, as in \cref{fig6}, panel a, and far more clearly than in \cref{fig6} b, where they were buried in the dairy cluster. The dairy products are arranged along a broad arc from the dense group at the lower left, through the centre of the map, to the mature and extra-mature Cheddars, which occupy an arm of their own at the lower right. Two differences from the measured map should be noted, because they reflect the model rather than the cheeses. First, the mature and extra-mature Cheddars, which overlapped with P01--P03 in \cref{fig6}, panel a, are separated from them in \cref{fig6} c: the predicted curves for these two products retain some of the distinctiveness of their acoustic signatures (\cref{fig6} b) even though their measured curves are close to those of the plant slices. Second, the stiff dairy products (Emmental, reduced-fat Cheddar, Red Leicester) are drawn towards the centre of the map rather than its edges, and hard Gouda separates into a group of its own. These are the products with the largest prediction errors in \cref{fig5}, and the compression of the dairy extremes towards the centre is the regression towards the mean discussed in \cref{sec:pred}. The predicted-curve map is therefore best read as a texture map whose plant-based and Cheddar-type regions are reliable and whose soft and stiff dairy extremes are smoothed.

Taken together, the three panels make a central methodological point of this work. The raw acoustic data do not organise by texture (\cref{fig6} b); the measured curves do (\cref{fig6} a); and the learnt mapping between them transfers that organisation to the non-contact measurement (\cref{fig6} c). The result is a texture map on which a new sample can be placed from a one-minute non-contact scan, without destroying it. For plant-based formulation this is the intended tool. A prototype can be positioned relative to established dairy targets and to competitor products, and the unoccupied regions of the map, the soft, mozzarella-like corner and the very stiff, Emmental-like corner that no commercial plant-based product in this study reaches, identify the textures that plant-based formulation has yet to deliver. That the commercial plant products occupy a narrow region of the map agrees with compositional surveys showing that commercial plant-based cheeses are formulated from a similar set of starches, coconut oil and hydrocolloids, \cite{Grasso2021PlantBasedBlocks,Grossmann2021PlantBasedCheese}, and with the narrow functional range reported for such products, \cite{Dobson2023PlantBasedCheese,Dobson2025StructuralDifferences}. It also indicates where a rapid screening measurement would be most useful: the hardness of a high-protein plant-based cheese can be tuned across the full dairy range (\qtyrange{15}{90}{\newton}) through formulation and processing, \cite{Dobson2023PlantBasedCheese}, and a non-contact map of this kind allows each iteration to be compared with its dairy target without a destructive test.

\subsection{Practical implications and limitations} \label{sec:limits}

In practical terms the approach replaces a destructive test that requires cutting, stacking and compressing sub-samples with a measurement that needs no couplant, no sample preparation and no contact. Each \qtyproduct{20 x 20}{\milli\metre} scan took approximately \qty{1}{\minute}, inference from the trained network is effectively instantaneous, and the slice is left intact and available for other measurements (of the same slice during storage or after heating, for example) or for sale. These properties suit at-line quality control of sliced products and comparative screening during formulation, and the through-transmission geometry used here is the one already used for non-contact inspection and imaging of food products with air-coupled transducers, \cite{Gan2006NonContactUltrasound}, so the measurement is in principle compatible with conveyor implementation as part of the sensor-based monitoring envisaged for digitalised food manufacturing, \cite{Watson2021IntelligentSensors}. Because ultrasound is a mechanical wave, the measurement responds to the elastic properties that underlie texture rather than to surface colour, an advantage over optical routes to cheese hardness, \cite{Vasquez2018CheeseHardness} for products as visually diverse as annatto-coloured Red Leicester, smoked slices and carotene-coloured plant-based slices.

Several limitations qualify these results. Each product was represented by a single retail purchase measured within a few weeks of purchase, so batch-to-batch variation and the evolution of texture during ripening and storage are not captured, although both composition and storage conditions are known to affect the TPA parameters of sliced cheese, \cite{Zheng2016CheeseTexture} and ripening to change its ultrasonic properties, \cite{Benedito2000CheddarUltrasound,Crespo2020TortaDelCasar}. Generalisation was demonstrated to unseen slices of known products; performance on products absent from training is untested, and the mozzarella case shows that the network defaults to the majority behaviour when a signature is unfamiliar, so extending the product set and the number of batches, and adopting a leave-one-product-out evaluation, are the priorities for further work. The dataset of 116 slices is modest for a convolutional model and constrained the architecture to a compact network. The texture reference was a \qty{50}{\percent}-strain double-compression test on refrigerated samples, which characterises firmness and the loading and unloading response but not the melting and stretching behaviour in which commercial plant-based cheeses differ most from dairy, \cite{Dobson2023PlantBasedCheese,Dobson2025StructuralDifferences}; the non-contact format lends itself to measuring heated samples, which is a natural extension. Finally, the acoustic features used were limited to time-domain amplitude and time-of-flight maps at a single nominal frequency; broadband excitation and velocity-derived moduli, which have proved informative for the mechanical properties of plant-based meat analogues, \cite{Guillermic2023MeatAnalogues,Rangira2026PeaProteinUltrasound}, and the full received waveform remain to be exploited.

The mechanical openings in the Cheddar, Gouda and Emmental slices deserve a separate comment, because they bear on both the dataset and any future implementation. They were avoided where possible when scan positions were chosen but were not recorded systematically, so their effect on the acoustic images and on the prediction errors cannot be quantified here. Three responses are available for future work. The presence and position of openings in each scanned area could be recorded, for which a photograph of the slice alongside the C-scan is sufficient; the C-scan itself could be used to detect them, since an opening produces a compact group of pixels whose amplitude departs strongly from the slice median, and the affected pixels could then be masked or passed to the network as an additional channel; and, because openings are a genuine feature of these products, a larger dataset would allow the network to learn to disregard them rather than requiring their removal. The last of these matters most for at-line use, where scan positions cannot be chosen by hand and the measurement must be robust to whatever structure the slice presents.

\section{Conclusions}

This study shows that non-contact ultrasonic imaging, interpreted with machine learning, can measure the mechanical texture of both dairy and plant-based cheese without contact, couplant or sample destruction. It also gives new insight into how the two categories differ.
On the science, the work makes three advances. First, air-coupled C-scans spatially resolve the internal organisation of cheese. Ripened dairy cheeses show millimetre-scale acoustic domains consistent with protein-rich curd junctions and local variation in fat, moisture and mineral content. Commercial plant-based slices are acoustically uniform, as expected from a homogeneous starch and coconut-oil gel. This structural distinction has previously been seen only with destructive or synchrotron-based methods. Second, the texture gap between plant-based and dairy cheese is more precisely defined here than as a simple difference in softness. Average firmness was similar across the two categories. The plant-based products, however, occupied a narrow central band of the dairy range, were far more uniform, and failed by brittle fracture with strain-softening rather than by the ductile, largely plastic yielding of Cheddar. Third, scalar ultrasonic parameters such as velocity and attenuation do not calibrate for texture across structurally diverse products, but the spatial information in the full acoustic image does. A single compact convolutional network mapped two-channel C-scans to complete loading and unloading stress–strain curves for thirteen products, including four starch-based gels, without product-specific calibration. The whole curve captures hardness, yield or strain-softening behaviour and elastic recovery together. Prediction errors were largest for products whose mechanical behaviour lies at the extremes of the dataset, but whose acoustic signatures are not distinctive, notably mozzarella, Emmental and hard Gouda. This identifies where further data are most needed.
For industry, the approach turns a destructive laboratory test into a one-minute, couplant-free measurement that leaves the product intact. Because the measurement is mechanical rather than optical, it does not depend on colour, smoking or surface finish, and it's through-transmission geometry is compatible with conveyor-based inspection. Two applications follow directly. During plant-based product development, a prototype can be placed on a texture map alongside its dairy target and competitor products after each formulation iteration, without cutting, stacking and compressing samples. The map produced here already shows where commercial products sit: the starch and coconut-oil slices are Cheddar-like and Violife is Emmental-like. It also shows that soft, mozzarella-like and very stiff textures are not yet delivered by any of the plant-based products tested. In manufacturing, the same measurement could support at-line monitoring of slice-to-slice consistency, where the higher natural variability of dairy products is greatest.
Realising these applications will require a larger dataset spanning multiple batches and storage times, evaluation on products absent from training, robust handling of eyes and mechanical openings, and extension to heated samples to capture the melt and stretch behaviour in which plant-based cheeses differ most from dairy. With these steps, non-contact ultrasonic texture mapping could become a practical tool for accelerating the development of plant-based cheeses that match the full range of dairy textures and further developed for other food products. 

\section*{Acknowledgements}
Nicholas Watson and Thomas Hazlehurst would like to thank the UK National Alternative Protein Innovation Centre (NAPIC), which is an Innovation and Knowledge Centre funded by the Biotechnology and Biological Sciences Research Council (BBSRC) and Innovate UK (Grant Ref: BB/Z516119/1) for supporting this work.

% Use the style of numbering in square brackets.
% If nothing is used, default style will be taken.
%\begin{enumerate}[a)]
%\item 
%\item 
%\item 
%\end{enumerate}  

%\begin{itemize}
%\item 
%\item 
%\item 
%\end{itemize}  

%\begin{description}
%\item[]
%\item[] 
%\item[] 
%\end{description}  

% % Figure
% \begin{figure}%[]
%   \centering
% %    \includegraphics{}
%     \caption{}\label{fig1}
% \end{figure}

% \begin{table}%[]
% \begin{tabular*}{\tblwidth}{@{}LL@{}}
% \toprule
% \midrule
%  & \\
%  & \\
%  & \\
%  & \\
% \bottomrule
% \end{tabular*}
% \end{table}

% Uncomment and use as the case may be
%\begin{theorem} 
%\end{theorem}

% Uncomment and use as the case may be
%\begin{lemma} 
%\end{lemma}


% \section{}\label{}

\printcredits

\bibliographystyle{cas-model2-names}

\bibliography{cas-refs}



\end{document}